\documentclass{elsarticle}

\usepackage[utf8]{inputenc}
\usepackage{graphicx}
\usepackage{amssymb}

\usepackage[margin=2.0cm]{geometry}%
\usepackage{lipsum}
\usepackage{xcolor}
\usepackage{tabularx}
\biboptions{sort&compress}

\journal{NPA}

\begin{document}
\begin{frontmatter}


\title{Energy dependence of the optical potential of the weakly bound $^{9}$Be projectile on the $^{197}$Au target}


 \author[label1,label2]{F. Gollan}
  \author[label1]{D. Abriola}
  \author[label1,label2]{A. Arazi}
  \author[label1,label2]{M. A. Cardona}
  \author[label1]{E. de Barbar\'a}
  \author[label1,label2]{D. Hojman}
  \author[label4,label5]{R.M.~Id~Betan}
  \author[label1]{G.V. Mart\'i}
  \author[label1,label2]{A.J. Pacheco}
   \author[label1,label2]{D. Rodrigues}
    \author[label1]{M. Togneri}


 \address[label1]{Laboratorio TANDAR, Comisi\'on Nacional de Energ\'{\i}a At\'omica, BKNA1650 San Mart\'{\i}n, Argentina}
 \address[label2]{Consejo Nacional de Investigaciones Cient\'{\i}ficas y T\'ecnicas,  C1425FQB Buenos Aires, Argentina}
 \address[label4]{Instituto de F\'isica de Rosario (CONICET-UNR), Bv. 27 de Febrero 210 bis, S2000EZP Rosario}
 \address[label5]{Facultad de Ciencias Exactas, Ingenier\'ia y Agrimensura (UNR), Av. Pellegrini 250, S2000BTP Rosario, Argentina}

\begin{abstract}
In this work we measured elastic and inelastic angular distributions of the weakly bound $^{9}$Be projectile on the $^{197}$Au target at several bombarding energies from 84\% up to 140\% of the Coulomb barrier. The elastic angular distributions were analyzed using a phenomenological Woods-Saxon potential and a double folding S\~ao Paulo potential and the energy dependence was extracted. Angular distributions from two inelastic peaks  were compared with coupled channel calculations using reduced  transition probabilities available in the literature.  
The energy dependence of the two  interaction potential models show a similar trend in the region of the Coulomb barrier. Dispersion relation calculation demonstrates the presence of  the breakup threshold anomaly proposed for weakly bound systems.  
\end{abstract}


\end{frontmatter}


\section{Introduction}
\label{S:Introduccion}


Reaction mechanisms involving weakly bound projectiles at energies around the Coulomb barrier have been a matter of extensive theoretical and experimental study in the area of nuclear reactions. One of the main features of these nuclei, both stable ($^{9}$Be, $^{6}$Li, $^{7}$Li) and unstable ($^6$He, $^{7}$Be, $^{8}$Li, $^{8}$B, etc.), is the low threshold energy for breakup into their cluster constituents~\cite{canto2006}. This process which is triggered mainly by Coulomb (nuclear) interaction when scattering on a heavy (light) target nucleus, radically modifies the dynamics of the collision with respect to tightly bound nuclei. Therefore, the fully understanding of the influence of these channels on different reaction mechanisms is crucial to validate  theoretical models of nuclear interaction capable of describing their peculiar structure, and also to understand the role of these nuclei  in astrophysics~\cite{thompson2009}.

It is a well established fact that  the real and imaginary parts of the optical potential obtained from the elastic scattering for the tightly bound systems show a distinctive energy behavior around the Coulomb barrier, a phenomenon referred to as threshold anomaly (TA)~\cite{Mahaux1985,satchler1991}. The imaginary part of the optical potential shows a rapid decrease due to the closing of inelastic channels, while the real part peaks in strength with a bell-shaped maximum in the same energy region. The behavior of the real  optical potential is ascribed mainly to the coupling of several nonelastic channels to the elastic channel that produce an attractive dynamic polarization potential. This phenomenon is a direct consequence of the dispersion relation~\cite{Mahaux1985,mahaux1986,satchler1991,Nagarajan1985} imposes  causality~\cite{Toll1956, Mahaux1985,mahaux1986,satchler1991} in the scattering of heavy ion systems.

In opposition, for systems where at least one of the reaction partners is a weakly bound nucleus, the energy behavior of the optical potential obtained from elastic scattering  analysis have shown different results ~\cite{JFN2007,Biswas2008,Deshmukh2011,Maciel1999,Hussein2006,gomezcamacho2010,Fimiani2012,Figueira2010,Souza2007,keeley1994,camacho2007,gomes2009,lubian2001,figueira2006,oliveira2011,gollan2018,Arazi2018}. As a consequence of the low binding energy, the breakup channel remains open, even at energies below the Coulomb barrier~\cite{Hinde2002}. Different results have suggested that the effect of coupling to the breakup channel could produce a repulsive polarization potential that exceeds the attractive terms arising from coupling to bound states and, therefore, inhibit the usual TA~\cite{Sakuragi1987,keeley1994}. Further works for weakly bound systems have shown the existence of a different kind of anomaly~\cite{Maciel1999,Signorini2000,pakou2003,pakou2004}, later referred to as breakup threshold anomaly (BTA)~\cite{Hussein2006,Hussein2007}. In this case, the coupling to the breakup channel could produce an increase of the imaginary potential as the energy decreases  below the Coulomb barrier. Due to the dispersion relation~\cite{pakou2004}, this implies the appearance of a  repulsive polarization potential that decreases the strength of the real part of the optical potential.


As it was previously asserted, the results for the elastic scattering of the stable weakly bound nuclei  $^{6,7}$Li, $^{9}$Be have produced different outcomes. The results for $^{6}$Li on $^{58}$Ni~\cite{gomezcamacho2010}, $^{80}$Se~\cite{Fimiani2012}, $^{144}$Sm~\cite{Figueira2010}, $^{208}$Pb~\cite{Hussein2006} and also for the $^{9}$Be~+~$^{80}$Se~\cite{gollan2018} and $^{9}$Be~+~$^{64}$Zn~\cite{Gomezcamacho2007} are consistent with the presence of the BTA. On the other hand, the data for the elastic scattering of the $^{7}$Li projectile on $^{59}$Co~\cite{Souza2007}, $^{80}$Se~\cite{Fimiani2012}, $^{138}$Ba~\cite{Maciel1999}, $^{144}$Sm~\cite{Figueira2010} and $^{208}$Pb~\cite{keeley1994} show the presence of the usual TA. Finally, no conclusions could be extracted regarding any of these anomalies for the cases of $^{6}$Li on $^{27}$Al~\cite{JFN2007}, $^{64}$Ni~\cite{Biswas2008}, $^{112,116}$Sn~\cite{Deshmukh2011}, $^{138}$Ba~\cite{Maciel1999}, $^{7}$Li~+~$^{28}$Si~\cite{pakou2004} and $^{9}$Be on $^{12}$C~\cite{oliveira2011}, $^{27}$Al~\cite{gomes2004}, $^{120}$Sn~\cite{Arazi2018} and $^{144}$Sm~\cite{gomes2009}. These results expose the necessity for further experimental data, especially for energies close to the barrier, where the elastic scattering is dominated by the Coulomb interaction.  In the same manner, rigorous criteria to determine statistical and systematic error of the experimental data is important to determine the behavior of the extracted interaction potentials as a function of energy, mainly at energies below the Coulomb barrier~\cite{Abriola2015}.


In the present work we measured elastic and inelastic angular distributions of the weakly bound projectile $^{9}$Be from the odd mass $^{197}$Au target at energies 34~$\leq~E_{lab}~\leq$~48~MeV. The $^{9}$Be structure could be thought as a three-body $\alpha + \alpha + n$ ($S_n=1.57$~MeV), where no two constituents alone can form a bound cluster, the so called Borromean structure~\cite{Parkar2013}, and may undergo a prompt breakup into the three fragments. On the other hand, 
 $^{9}$Be has a low neutron separation energy ($S_n=1.67$~MeV) which favours the n-transfer process leading to the unstable $^{8}\textrm{Be}$  with a half life of 0.07~fs. The latter breaks up into two $\alpha$ particles~\cite{Rafiei2010} or through $^{9}\textrm{Be}\rightarrow^{5}\textrm{He}+\alpha$ with $S_\alpha=2.47$~MeV.  The target $^{197}$Au was chosen due to its relatively high mass which enhances the Coulomb breakup probability. 


One of the main ambiguities to conclusively establish whether the BTA is present for a given system relies on the different models used to describe the experimental elastic scattering data. In this work, two different potentials were used, the phenomenological Woods-Saxon potential~\cite{Maciel1999,gomes2004} and the microscopic double-folding S\~ao Paulo potential~\cite{gomes2004,Gomes2005}. This aims to assess to which extent the physical conclusions are model independent. The inelastic distributions due to target nucleus excitation were analyzed through coupled-channel (CC) calculations with matrix elements that explicitly include the Coulomb and nuclear deformation from coupling of the ground state to (and between) the excited states experimentally observed and a bare Woods-Saxon potential that reproduce the projectile-target nuclei interaction in the  absence of coupling to internal degrees of freedom. As stated before, the fundamental postulate underlying the energy dependence of the nuclear potentials is the dispersion relation. Its fulfilment was verified using a statistical method on random sampling to analyze the correlation between the real and imaginary parts of the optical potential.

The present article is organized as follows. In Sec.~\ref{experimental}, the experimental setup is addressed and experimental elastic and inelastic angular distributions are presented. In Sec.~\ref{resultados} the elastic scattering angular distributions are calculated with the Woods-Saxon and S\~ao Paulo potentials. Inelastic angular distributions are analyzed with CC calculations. The energy dependence of the nuclear potentials were evaluated through the dispersion relation.  Finally, in Sec.~\ref{conclusiones} summary and general conclusions are presented. 
	
\section{Experimental setup}
\label{experimental}

Angular distributions for the $^{9}$Be~+~$^{197}$Au were measured at the TANDAR Laboratory in Buenos Aires, Argentina. $^{9}$Be beams were delivered by the 20UD tandem accelerator at twelve bombarding energies: 34, 35, 36, 37, 38, 39, 40, 41, 42, 44, 46 and 48 MeV. The nominal value of the Coulomb barrier was estimated in 40.5~MeV (38.7~MeV in c.m.) by empirical models~\cite{Wilczynski1975}. The $^{197}$Au target consists of a thin self-supporting foil of 250~$\mu$g/cm$^2$ placed at the center of a 70~cm diameter scattering chamber. The target angle was set at +40$^{\circ}$ (-40$^{\circ}$) relative to the beam direction for the measurements of forward (backward) angles. For the data analysis the beam energy was corrected for the energy lost (about 60~keV for $34~\leq~E_{lab}~\leq~41$~MeV and 50~keV for $E_{lab}~\geq~42$~MeV) assuming that the reaction takes place in the middle of the target. The beam was defined by two rectangular collimators (7~mm $\times$~9~mm) located 51~cm and 156~cm upstream the scattering chamber. A third circular collimator with a diameter of 7.6~mm was located at the entrance of the scattering chamber.  

\begin{figure}
	\centering
	\includegraphics[width=0.65\linewidth]{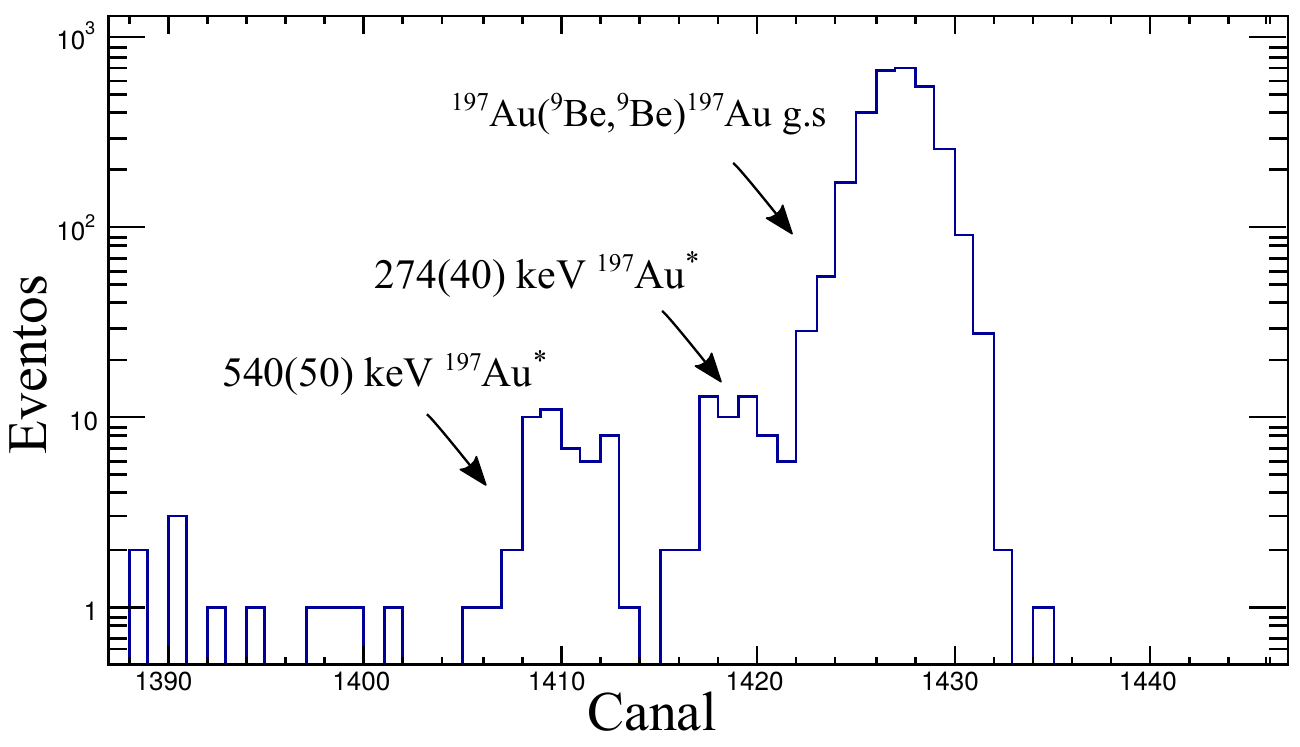}  
	\caption{Energy spectrum of the reaction products for the $^{9}$Be + $^{197}$Au system, obtained at $E_{\rm lab} = 48$~MeV, $\theta_{\rm lab} = 67.6^{\circ}$. Different channels correspond to elastic and inelastic scattering corresponding to  the  $^{197}$Au excited states.} 
	\label{fig:espectro}
\end{figure}

The detection system consisted in an array of eight silicon surface-barrier detectors, placed on a rotatable plate at the bottom of the scattering chamber, with an angular separation of 5$^{\circ}$ between adjacent detectors. Two additional  silicon detectors with an angular separation of 8$^{\circ}$ were mounted on a top rotatable plate in the scattering chamber. The detector setup was previously used by our group to measure the elastic scattering angular distributions of the $^{9}$Be~+~$^{80}$Se system~\cite{gollan2018}. The energy resolution of the detectors (from 0.2 to 0.5\%)  allowed us to separate two inelastic-excitation peaks with excitation energies of 274(40)~keV and 540(50)~keV from the elastic peak, as it is shown in Fig.~\ref{fig:espectro}. Given that there are no excited states of $^{9}$Be above the breakup threshold energy, all inelastically  scattered nuclei are the result of excitations of the $^{197}$Au target. For the low energy  peak, the states that can contribute are the $(268.8~\rm{keV},3/2^{+})$ and $(279.0~\rm{keV}, 5/2^{+})$. For the high  energy peak, the states $(502.5~\rm{keV},5/2^{+})$, $(547.5~\rm{keV},7/2^{+})$ and the 583.0~keV (unknown spin parity) are compatible with the observations. Since the energy separation between different states is  significantly lower than the resolution of the detectors (estimated in 100~keV), it was not possible to separate them individually. Following the referred works of~\cite{Sharma1970,Nelson1971, Bolotin1979}, we assume that the inelastic peaks correspond mainly to the 279.0 and 547.5~keV states. On the other hand, background events, mostly alpha particles from the projectile breakup~\cite{Arazi2018}, have lower energy and hence produce no interference in the region of the elastic peak.

The detectors were collimated by rectangular slits, defining its angular acceptance. Solid angles ranged from 0.1 msr for the most forward detector to 0.6 msr for the most backward one. This ensured a similar counting rate in all detectors. The normalization of the elastic and inelastic cross sections were carried out by using two silicon monitor detectors, fixed at +16$^{\circ}$ and -16$^{\circ}$ where the scattering is pure Rutherford. Through this method, the differential cross section for the \textit{i}th detector ($i=1-10$) at an angular position $\theta_{\rm{Det}}^{(i)}$ and normalized with the \textit{j}th monitor ($j=1,2$) is determined from the expression  
\begin{equation}\label{eq:se_met_monitor}
\frac{d\sigma}{d\Omega}(\theta_{\rm{Det}}^{(i)})=\frac{d\sigma_{\rm{Ruth}}}{d\Omega}(\theta_{\rm{Mon}}^{(j)})\frac{\Omega_{\rm{Mon}}^{(j)}}{\Omega_{\rm{Det}}^{(i)}}\frac{N_{\rm{Det}}^{(i)}}{N_{\rm{Mon}}^{(j)}}\frac{J(\theta_{\rm{Det}}^{(i)})}{J(\theta_{\rm{Mon}}^{(j)})},  
\end{equation}
where $N_{\rm{Det}}^{(i)}$  and $N_{\rm{Mon}}^{(j)}$ are the numbers of events, $J(\theta_{\rm{Det}}^{(i)})$  and $J(\theta_{\rm{Mon}}^{(j)})$ are the Jacobian factors for the laboratory to center-of-mass transformation and $\Omega_{\rm{Mon}}^{(j)}/\Omega_{\rm{Det}}^{(i)}$ is the solid angle ratio between the $j^{th}$ monitor and the $i^{th}$ detector. The use of the two monitors provided us with corroboration of the results. To determine the solid angle ratios, several elastic scattering cross section of $^{9}$Be and $^{16}$O in $^{197}$Au were measured for energies well below their corresponding Coulomb barriers. An alternative and independent method to corroborate the normalization was supplied by the charge collected by a Faraday cup at the end of the beam line, 6~m away from the scattering chamber.     

\begin{figure}
	\centering
	\includegraphics[width=0.85\linewidth]{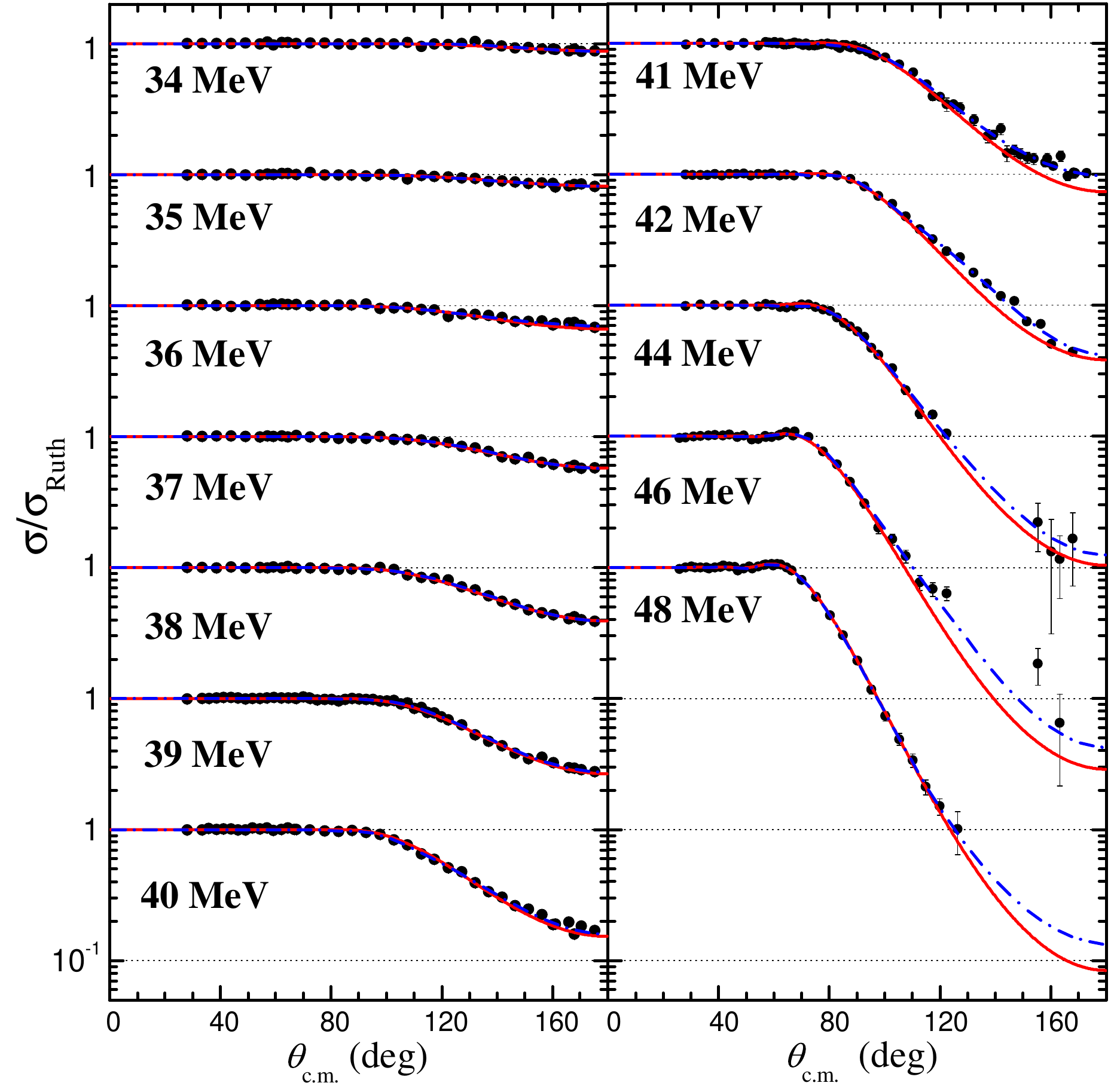}
	\caption{(Color online) Experimental elastic scattering differential cross sections normalized to the Rutherford cross section for the $^{9}$Be~+~$^{197}$Au system (open circles) and fits from optical model calculations. The dash-dot blue lines correspond to the phenomenological  Woods-Saxon energy-dependent with variable geometry potential. The full red lines correspond to the double folding S\~ao Paulo potential.} 
	\label{fig:DA_elasticas}
\end{figure}

A summary of the experimental elastic scattering angular distributions  for the $^{9}$Be~+~$^{197}$Au system normalized to the Rutherford cross section is displayed in Fig.~\ref{fig:DA_elasticas}. The overall uncertainties were  estimated in the range of 2-6\%, except for the higher energies and backward angles for which they raised up to 10\%. The bars reflect the statistical uncertainties as well as those from stemmed from the solid angle calculation. In Fig~\ref{fig:inelasticos} the experimental inelastic angular distributions for the two analyzed peaks are displayed (inelastic peaks could not be resolved for $E_{lab}=41$ MeV). From the two figures one can see that the angular distribution for the first inelastic state spans up to approximately $90^{\circ}$. This effect is due to the widening of the elastic peak at backward angles, that make impossible to identify the inelastic peak closer to it. 

\section{Data analysis and results}
\label{resultados}
\subsection{Optical model analysis} \label{modelo_optico}

The experimental elastic scattering data were analyzed in the framework of the optical model. The first potential used was the phenomenological Woods-Saxon potential~\cite{woods1954} defined as 
\begin{equation}\label{eq:ws_pot1}
U(r)=V_{C}(r)-Vf(r,r_{0},a)-iW_{i}[f(r,r_{i0},a_{i})]^2-iW_{si}g(r,r_{si0},a_{si}).
\end{equation}
The first term corresponds to the Coulomb potential of a charged sphere with radius $R_C=1.2(A_{p}^{1/3}+A_{t}^{1/3})$ where  $A_{p}$ and $A_{t}$ are the projectile and target mass numbers. The second term represents the real part of the nuclear potential $V(r)$ where the radial distribution \textit{f} is the Woods-Saxon form factor 
\begin{equation}\label{eq:WSfactor}
f(r,r_{0},a)=\frac{1}{1+e^{[r-R(r_{0})]/a}}.
\end{equation}
Here, $R(r_0)$ is the  radius of the potential, \textit{a} the diffuseness (geometrical parameters) and \textit{V} the depth of the real potential. It is usual to represent the radius as a function of the reduced radius $r_0$ with $R(r_0)=r_0(A_{p}^{1/3}+A_{t}^{1/3})$. The same expressions are obtained for the reduced radii of the imaginary (absortive) potential, for which a volume and a surface contributions are assumed. The volume contribution $W_{vol}(r)$ (third term) is proportional to the square of the Woods-Saxon form factor with a volume depth $W_i$. It simulates the incoming wave boundary condition and accounts for fusion~\cite{Abriola1992}. The surface contribution (fourth term) has a form factor \textit{g} which is proportional to the derivative of the Woods-Saxon shape
\begin{equation}\label{eq:WSfactorsup}
g(r,r_{si0},a_{si}) =-4 a_{si}\frac{df}{dr}(r,r_{si0},a_{si}),
\end{equation}
multiplied by the surface depth $W_{si}$, and represents the absorption due to the peripheral reactions. 

\begin{figure}[!tbp]
	\centering
	\begin{minipage}[b]{0.503\textwidth}
		\includegraphics[width=\textwidth]{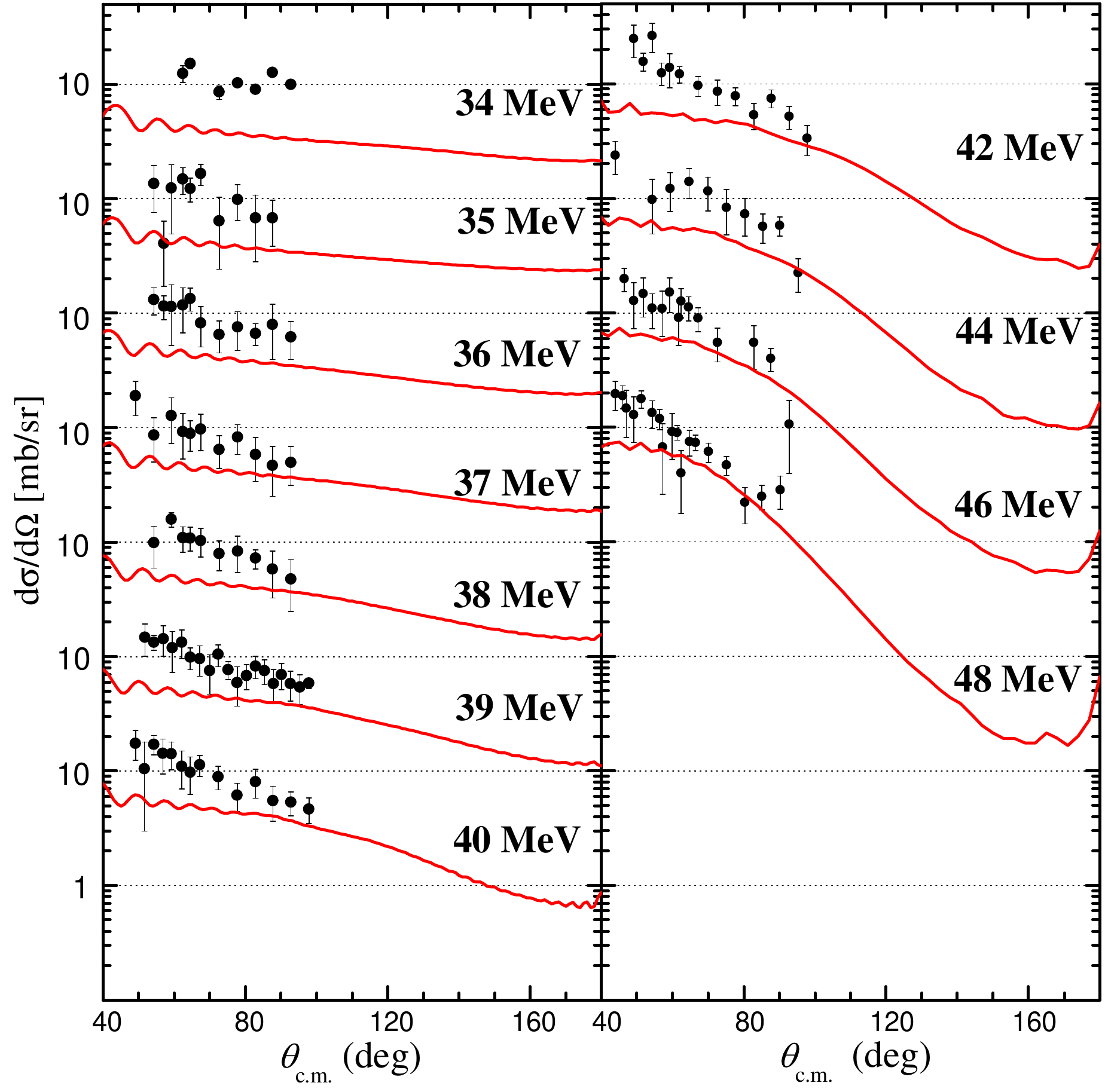}
	\end{minipage}
	\begin{minipage}[b]{0.48\textwidth}
		\includegraphics[width=\textwidth]{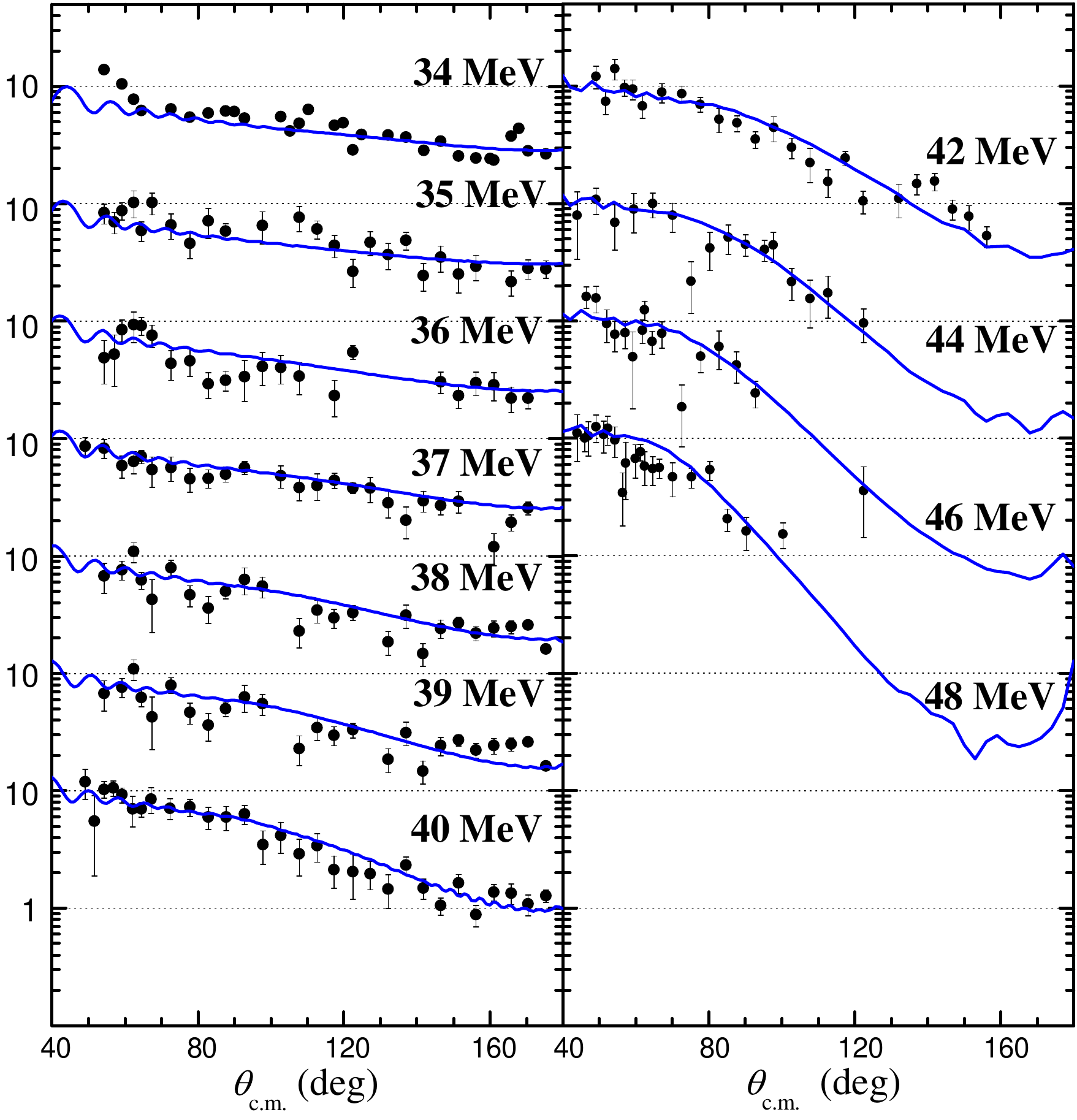}
	\end{minipage}
	\caption{Experimental inelastic scattering angular distributions for the  274(40)~keV and 540(50) keV excited states of the $^{197}$Au target. Full red and blue lines correspond to coupled channel calculations  performed for energy with the EDVG potential obtained in Sec.~\ref{modelo_optico} considering the $(279.0~\rm{keV}, 5/2^{+})$ (left figure) and $(547.5~\rm{keV},7/2^{+})$ (right figure) states of the target.}
	\label{fig:inelasticos}
\end{figure}

The interaction parameters of the nuclear potential were calculated under three different conditions: First, we searched for a global geometry by fitting all angular distributions with an energy-independent potential (EI); Second, we used an energy-dependent potential with fixed geometric parameters (EDFG); and last, an energy-dependent potential with variable geometric parameters (EDVG) was used. In all cases the imaginary volume parameters were fixed with values of $W_i=10$~MeV, $r_0=1.0$~fm, and $a_i=0.25$~fm. Since these geometrical values are much lower than the corresponding real and surface values, the volume potential plays no significant role in the outer nuclear region.
 The EI potential, defined by  real and imaginary surface parameters $V$, $r_0$, $a$, $W_{si}$, $r_{si0}$ and $a_{si}$, was obtained choosing starting values for these parameters from a  6-dimensional grid. From each set of values, the code PTOLEMY~\cite{macfarlane1978} was used to  fit all experimental data (all angles and energies) simultaneously. 
 As result of this simultaneous best fit procedure we obtained the following optical potential parameters: $V=27.4$~MeV, $r_0=1.306$~fm, $a=0.468$~fm, $W_{si}=2.40$~MeV, $r_{si0}=1.331$, and $a_{si}=0.648$~fm. The global estimator of the fit, defined as $\sum_{i=1}^{n} \chi^{2}_{i}/(\sum_{i=1}^{n}N_{i}-p)$, resulted equal to 1.244, where $p=6$ is the number of  adjusted parameters and $N_i$ is the number of experimental points for each energy ($i=1-12$). The potentials EDFG and EDVG were calculated with the code SFRESCO~\cite{thompson1988}. This code was selected because it was later used for the coupled channel calculations. The consistency of the results obtained with PTOLEMY was checked by performing individual calculations using SFRESCO. The EDFG potential was obtained keeping  all the geometrical values (reduced radii and diffuseness) fixed to the EI case and fitting the real and surface potential depths ($V$ and $W_{si}$) as a function of the energy. The results are presented in Table~\ref{tabla_fresco_DEGF}.
\begin{table*}[t]
	\centering
	\caption{\label{tabla_fresco_DEGF} Optical parameters for the Woods-Saxon energy-dependent with fixed geometry (EDFG) potential obtained from the fit performed on the experimental elastic scattering angular distributions for the $^{9}$Be~+~$^{197}$Au system. The global estimator of the fit $\sum_{k=1}^{n} \chi^{2}_{k}/(\sum_{k=1}^{n}N_{k}-p)$ is equal to  0.685, where $p=2$ is the number of adjusted parameters and \textit{k} stands for each energy.}
	\resizebox{\textwidth}{!}{%
	\begin{tabular}{ccccccccccccc} 
		\hline\hline
		$E_{\rm c.m.}$ & $V$ & $r_{0}$ &  a  & $W_{i}$ & $r_{i0}$ & $a_{i}$ & $W_{si}$ & $r_{si0}$ & $a_{si}$ & $[G(E)]_V$ & $[G(E)]_W$ & $\chi^{2}/\nu$  \\
		
		\footnotesize{(MeV)}          & \footnotesize{(MeV)}& \footnotesize{(fm)}    &\footnotesize{(fm)} & \footnotesize{(MeV)}   &\footnotesize{ (fm) }     & \footnotesize{(fm)}    & \footnotesize{(MeV)}   &\footnotesize{ (fm) }     & \footnotesize{(fm)}     &\footnotesize{(MeV fm$^3$)}&\footnotesize{(MeV fm$^3$)}\\
		[0.2ex] 
		
		\hline
		32.5  & 220.4  & \mbox{\vline height 1ex depth 1ex} & \mbox{\vline height 1ex depth 1ex} & \mbox{\vline height 1ex depth 1ex} & \mbox{\vline height 1ex depth 1ex} & \mbox{\vline height 1ex depth 1ex} & 1.37 & \mbox{\vline height 1ex depth 1ex} & \mbox{\vline height 1ex depth 1ex}   &  9.5   &  0.5   & 0.460 \\ 
		33.5  & 1.6    & \vline & \vline & \vline & \vline & \vline & 4.49 & \vline & \vline   &  0.1   &  1.5   & 0.350 \\ 
		34.4  & -29.0 & \vline & \vline & \vline & \vline & \vline & 5.80 & \vline & \vline    &  -1.3  &  1.9   & 0.533 \\ 
		35.4  & 20.0   & \vline & \vline & \vline & \vline & \vline & 4.10 & \vline & \vline   &  0.9   &  1.4   & 0.526 \\ 
		36.3  & 34.8   & \vline & \vline & \vline & \vline & \vline & 3.94 & \vline & \vline   &  1.5   &  1.3   & 0.460 \\ 
		37.3  & 36.4   & 1.306  & 0.468  & 10.0   & 1.0    & 0.25 &   2.97 & 1.331   & 0.648   &  1.6   &  1.0   & 0.612 \\ 
		38.3  & 25.6   & \vline & \vline & \vline & \vline & \vline & 3.90 & \vline & \vline   &  1.1   &  1.3   & 1.111 \\ 
		39.2  & 22.1   & \vline & \vline & \vline & \vline & \vline & 4.27 & \vline & \vline   &  1.0   &  1.4   & 0.661 \\ 
		40.2  & 33.0   & \vline & \vline & \vline & \vline & \vline & 3.32 & \vline & \vline   &  1.4   &  1.1   & 1.565 \\ 
		42.1  & 31.8   & \vline & \vline & \vline & \vline & \vline & 3.93 & \vline & \vline   &  1.4   &  1.3   & 0.636 \\ 
		44.0  & 32.7   & \vline & \vline & \vline & \vline & \vline & 3.32 & \vline & \vline   &  1.4   &  1.1   & 1.467 \\ 
		45.9  & 43.0   & \mbox{\vline height 2ex depth -.5ex} & \mbox{\vline height 2ex depth -.5ex} & \mbox{\vline height 2ex depth -.5ex} & \mbox{\vline height 2ex depth -.5ex} & \mbox{\vline height 2ex depth -.5ex} & 3.41 & \mbox{\vline height 2ex depth -.5ex} & \mbox{\vline height 2ex depth -.5ex}   &  1.9   &  1.1   & 0.327  \\

		\hline\hline
	\end{tabular}}
\end{table*}
As can be seen, the adjustments showed an improvement over the previous case. With this consideration in mind, we performed a simultaneous fit of the potential depths and geometrical parameters (EDVG potential) with the condition that the latter do not vary significantly from those obtained in the energy independent case. In all cases, the uncertainties were calculated with a 68\% confidence interval by varying the parameters until the $\chi^2$ is increased respect to the minimum $\chi_{0}^2$ by $\Delta \chi^2~\chi_0^2/\nu$. The $\Delta \chi^{2}$ factor is equal to 2.3 and 7.04 for two and six parameters, respectively~\cite{Abriola2015,Press2007}. The results for the EDVG potential are displayed in Table~\ref{tabla_fresco_DEGV}.
\begin{table*}[t]
	\centering
	\caption{\label{tabla_fresco_DEGV} Optical parameters for the Woods-Saxon energy-dependent with variable geometry (EDVG) potential obtained from the fit performed on the experimental elastic scattering angular distributions for the $^{9}$Be~+~$^{197}$Au system.  The global estimator of the fit $\sum_{k=1}^{n} \chi^{2}_{k}/(\sum_{k=1}^{n}N_{k}-p)$ is equal to  0.575, where $p=6$ is the number of adjusted parameters and \textit{k} stands for each energy.}
	\resizebox{\textwidth}{!}{%
	\begin{tabular}{ccccccccccccc}
		\hline\hline
		$E_{\rm c.m.}$ & $V$ & $r_{0}$ & a & $W_{i}$ & $r_{i0}$ & $a_{i}$ & $W_{si}$ & $r_{si0}$ & $a_{si}$ & $[G(E)]_V$ & $[G(E)]_W$ &  $\chi^{2}/\nu$  \\ 
		\footnotesize{(MeV)}          & \footnotesize{(MeV)}& \footnotesize{(fm)}    &\footnotesize{(fm)} & \footnotesize{(MeV)}   &\footnotesize{ (fm) }     & \footnotesize{(fm)}    & \footnotesize{(MeV)}   &\footnotesize{ (fm) }    & \footnotesize{(fm)}     &\footnotesize{(MeV fm$^3$)}&\footnotesize{(MeV fm$^3$)}& \\
		[0.2ex] 
		\hline
		32.5 &  27.8 & 1.400  & 0.537   &  \mbox{\vline height 1ex depth 1ex} &  \mbox{\vline height 1ex depth 1ex} &  \mbox{\vline height 1ex depth 1ex} & 0.46  & 1.390 & 0.735   &   4.6    &   0.3   & 0.468   \\
		33.5 & 17.0  & 1.340  & 0.417   & \vline & \vline & \vline & 3.03  & 1.321  & 0.710  &   1.4    &   1.1   & 0.386    \\
		34.4 & -5.3 & 1.375  & 0.571  & \vline & \vline & \vline &  7.02  & 1.358 & 0.647    &   -0.7   &   2.9   &  0.597    \\
		35.4 & 12.9  & 1.360  & 0.463  & \vline & \vline & \vline & 3.30 & 1.360 & 0.609     &   1.1    &   1.3   & 0.495   \\
		36.3 & 29.7  & 1.350  & 0.402   & \vline & \vline & \vline & 2.59  & 1.320 & 0.731   &   1.9    &   1.0   & 0.503 \\
		37.3 & 30.1  & 1.340  & 0.416   &  10.0  &   1.0   & 0.25  & 2.00  & 1.380 & 0.571   &   1.7    &   0.9   & 0.537    \\
		38.3 & 22.3  & 1.332  & 0.443    & \vline & \vline & \vline & 2.07  & 1.362 & 0.686  &   1.3    &   0.9   & 1.131    \\
		39.2 &  18.6 & 1.339  & 0.431   & \vline & \vline & \vline & 3.53  & 1.310 & 0.710   &   1.1    &   1.1   & 0.552    \\
		40.2 & 21.2  & 1.329  & 0.443   & \vline & \vline & \vline & 2.47  & 1.384 & 0.613   &   1.1    &   1.2   & 0.924    \\
		42.1 & 30.4  & 1.330  & 0.409   & \vline & \vline & \vline & 2.23  & 1.355 & 0.710   &   1.5    &   1.0   & 0.672    \\
		44.0 & 28.4  & 1.340  & 0.434   & \vline & \vline & \vline & 3.90  & 1.349 & 0.570   &   1.7    &   1.3   & 1.298    \\
		45.9 & 38.8  & 1.333  & 0.425   & \mbox{\vline height 2ex depth -.5ex} & \mbox{\vline height 2ex depth -.5ex} & \mbox{\vline height 2ex depth -.5ex} & 2.94  & 1.319 & 0.690   &   2.1   &   1.0    & 0.358    \\ 
		
		\hline\hline
	\end{tabular}}
\end{table*} 
Although the global estimator indicates an improvement in the fit, the $\chi^2/\nu$ values ($\nu=N-p$ is the degree of freedom) do not exhibit a significant enhancement when compared to the EDFG case. 

The other potential considered in this work was the S\~ao Paulo potential~\cite{Chamon1997,Chamon2002,Alvarez2003}. The normalized version~\cite{Hussein2006} that we used to analyzed the experimental data is given by 
\begin{equation}
V_{SP}(r,E)=[N_R(E)+iN_I(E)]V_{LE}(r,E),
\end{equation}
where $N_R(E)$ and $N_I(E)$ are normalization factors of the real and imaginary parts of the potential which are energy dependent and $V_{LE}$ is the local equivalent potential. In the table~\ref{tabla_SPP} the adjusted values for the experimental data together with the volume integrals per nucleon are displayed.
\begin{table*}[t]
	\centering
	\caption{\label{tabla_SPP}Normalization factors $N_R$ and $N_I$ of the S\~ao Paulo potential (SPP) for the $^{9}$Be~+~$^{197}$Au system.  The volume integrals were evaluated at the $R_S$ radius. The global estimator of the fit $\sum_{k=1}^{n} \chi^{2}_{k}/(\sum_{k=1}^{n}N_{k}-p)$ is 0.760, where $p=2$ is the number of parameters and \textit{k} stands for each energy.}
	\begin{tabular}{cccccc} 
		\hline\hline
		$E_{\rm c.m.}$                 & $N_R$ & $N_I$ & $[G(E)_V]$   & $[G(E)_W]$  & $\chi^{2}/\nu$  \\
       \footnotesize{(MeV)}  &        &         &\footnotesize{(MeV fm$^3$)}&\footnotesize{(MeV fm$^3$)}& \\     
      [0.2ex] 
		\hline
		32.5 &  1.72   &   1.13   &  2.9  & 1.2    & 0.544  \\   
		33.5 &  0.25   &   1.50   &   0.3 &  1.9    & 0.357   \\  
		34.4 &  -0.43  &   2.02   &   -0.5&  2.6   &  0.536   \\ 
		35.4 &  0.70   &   1.28   &   0.6 &  1.8   & 0.581   \\  
		36.3 &  1.04    &   1.16  &   1.0 &  1.8       & 0.720    \\ 
		37.3 &  0.98    &   0.98  &   1.2 &  1.3      & 0.611   \\      
		38.3 &   0.92   &   1.07  &   1.1  & 1.5       & 1.284\\     
		39.2 &  0.82   &  1.21    &   1.0  & 1.5   & 0.757  \\   
		40.2 &  0.88   &  1.16    &   1.1  & 1.5   & 1.575  \\   
		42.1 &  0.93   &  1.26    &   1.2  & 1.6   &  0.765  \\  
		44.0 &  0.85   &  1.20    &   1.1  & 1.5   &  1.551  \\  
		45.9 &  1.04   &  1.22    &   1.3  & 1.5   & 0.384 \\    
		\hline\hline
	\end{tabular}
\end{table*} 
The $\chi^{2}/\nu$ values show that this potential correctly adjust the experimental data, although slightly poorer when compared to the Woods-Saxon EDVG potential.

\subsection{Inelastic scattering analysis} \label{analisis inelastico}

As it was shown in Fig.~\ref{fig:espectro}, two peaks corresponding to inelastic excitation were separated from the elastic scattering peak.
To describe the positive-parity inelastic states of the odd-mass-number gold isotopes in terms of collective states we used  an extended version of the particle-asymmetric rotor model~\cite{Vieu_1978,Hecht1962}. In this way, the coupling of the odd proton to an asymmetric rotating core gives rise to two rotational bands, namely $K=3/2$ g.s. and $K=1/2$, built on the first excited state located at and excitation energy $E_x=77$~keV.

We performed simple coupled channel (CC) calculations within the collective model with both Coulomb and nuclear deformation potentials. To carry out the calculation, coupling from g.s. to $j^{\pi}=5/2^{+}$ (279 KeV) and to $7/2^{+}$ (547 keV) and also between these two excited estates were considered, while reorientation terms were disregarded. We also assumed that both excited states belong to the $3/2^{+}$ band.
Within the rotational model, and assuming the same mass and charge distribution, the deformation parameters can be obtained by $\beta_{\lambda}=4\pi/3ZR_{0}^{\lambda-1}[B(E\lambda)\uparrow/e^{2}]^{1/2}$ where $R_0$ is the average radius and $\lambda$ stands for the transition multipolarity. The reduced transition probabilities were extracted from~\cite{Huang2005}. To simplify the analysis, we considered only the $E2$ component in $3/2^{+}_{1}(\rm{g.s.})\rightarrow 5/2^{+}_{1}$ and $5/2^{+}_{1}\rightarrow 7/2^{+}_{1}$ transitions. Finally, we used the Woods-Saxon EDVG potential obtained for each energy in Sec.~\ref{modelo_optico} for the projectile-target interaction in the absence of coupling to internal degrees of freedom. 

In Fig.~\ref{fig:inelasticos} the CC calculations are displayed in full lines. For the higher excited state, values are in good agreement (within the uncertainty bars) with the experimental data in a wide angular interval. This is a strong indication that the $7/2^{+}$ excited state is the predominant contribution. Instead,  the lower energy  state shows that CC are below the experimental data, especially for the lower bombardment energies. This difference is likely due to two effects: the proximity of the inelastic peak to the elastic one, whose  low-energy tail could have a contribution. This is supported by the fact that the difference increases as the incident energy diminishes and the elastic cross section increases in strength. On the other hand, the contribution of the $(268.8~\rm{keV}, 3/2^{+})$  which was not considered, could also have some contribution to the peak area. Nonetheless, the overall results show that this simple model provides a good description of the two excited states.


\subsection{Energy dependence} \label{dependencia con E}

It is evident from comparing the results in Tables~\ref{tabla_fresco_DEGF},~\ref{tabla_fresco_DEGV} and~\ref{tabla_SPP} the ambiguities between different potential conditions. For instance, the depth of the SPP potential was around 400~MeV for each energy, considerably larger than that of the shallow Woods-Saxon potential, usually around the tens of MeV.
 This situation can be overcome if the potentials are evaluated at the sensitivity radius $R_S$~\cite{satchler1991,Abriola1992,fulton1985} at each energy. The $R_S$ is the radial distance for which different potential with comparable goodness of fit (i.e., similar $\chi^2$) intersect each other or, equivalently where the elastic scattering data are most sensitive to the potential depths ($V$ and $W$) and the energy dependence is similar. For the real part of the potential, a set of diffuseness \textit{a} parameters were chosen around the optimal value (taken in steps of 0.05~fm) and kept fixed and adjust the $r_0$ and $V$ parameters to fit the experimental data to obtain $\chi^2/\nu$ closed to the optimum. Analogously, the sensitive radius of the imaginary part were calculated. This procedure was performed for each energy $E$  for the Woods-Saxon EDVG case (for details of the procedure and a  graphical example see~\cite{gollan2018}). The obtained radii fluctuated in the radial region $10.90~\rm{fm}\leq r \leq13.50~\rm{fm}$. A total of $N=16$ values were considered from the energies between 38 and 48~MeV. The average radius is $\bar{R}_S=(12.14\pm0.23)~\rm{fm}$. The uncertainty in this quantity has been estimated by means of a Student's \textit{t} of \textit{N}-1 degrees of freedom~\cite{figueira2006}. The interval corresponds to a confidence level of 70\%. 

\begin{figure}[!tbp]
	\centering
	\includegraphics[width=0.65\linewidth]{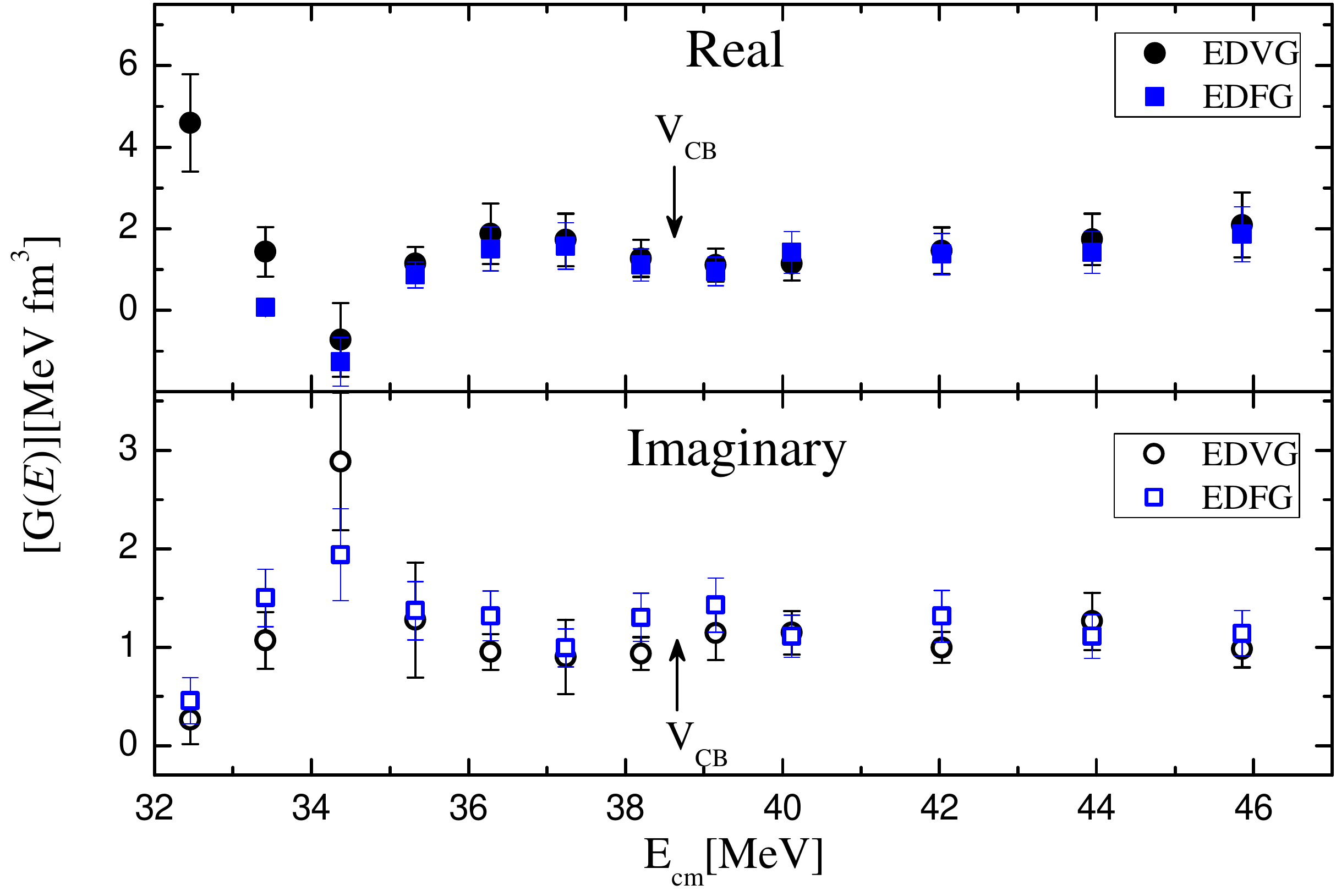}
		\caption{\label{fig:Ivol_WS}Volume integrals per nucleon $[G(E)]$ evaluated at the sensitivity radius calculated for the real (full symbols) and imaginary (open symbols) parts of the Woods-Saxon EDFG (blue squares) and EDVG (black circles) potentials analyzed in this work. Uncertainty bars include the  component of the sensitivity radius  $R_S$. The arrows indicate the value of the Coulomb barrier $V_{CB}$.}
\end{figure}

The fluctuation in the sensitivity radius implies that peripheral reactions cannot be narrowed to a single radial distance, but rather they are sensitive to a broader region. In consequence, the optical potentials are replaced by volume integrals per nucleon pair weighted by a Gaussian function~\cite{Brandan1993}. The volume integrals  are in the form of 
\begin{equation}\label{eq:IntegVol}
[G(\textit{E})]_{\psi} = \frac{1}{A_{p} \times A_{t}} \int{\psi(r,E)\phi(r)4\pi r^{2} dr}, 
\end{equation}   
where $\psi(r,E)$ stands for the real and imaginary parts (volume~+~surface) of the optical potential and $\phi(r)$ is the Gaussian probability density function centered at the sensitivity radius with a width $\sigma$. The volume integrals $[G(\textit{E})]_{V}$ and $[G(\textit{E})]_{W}$ are related trough a dispersion relation~\cite{mahaux1986,Mahaux1985}, in the same way as the optical potentials, as follows
\begin{equation}\label{eq:DR_Ivol}
[G(\textit{E})]_{V} = \frac{P}{\pi}  \int{\frac{[G(\textit{E'})]_{W}}{E'-E}}dE', 
\end{equation} 
where \textit{P} denotes the principal value of the Cauchy's integral. The sigma value was set to 0.7~fm to include the region of highest contribution to the potentials. The volume integral for the EDFG and EDVG cases were calculated and the results are displayed in Fig.~\ref{fig:Ivol_WS}. Uncertainties were calculated taking into account the contributions from the potential depths and the sensitivity radius.
It is clear from this figure that the energy dependence is the same for both cases. The  $[G(E)]_V$ and $[G(E)]_W$ values are  approximately constant for energies above the Coulomb barrier ($V_{CB}^{\rm c.m.}=38.7$~MeV) and hold up to $E_{\rm c.m.}=36.3$~MeV. From this value, the strength of $[G(E)]_V$ starts to decrease to a minimum around $E_{\rm c.m.}=34.4$~MeV and then it increases as the energy decreases. For $[G(E)]_W$ the behavior is the opposite in the same energy region. 
\begin{figure}
	\centering
	\includegraphics[width=0.65\linewidth]{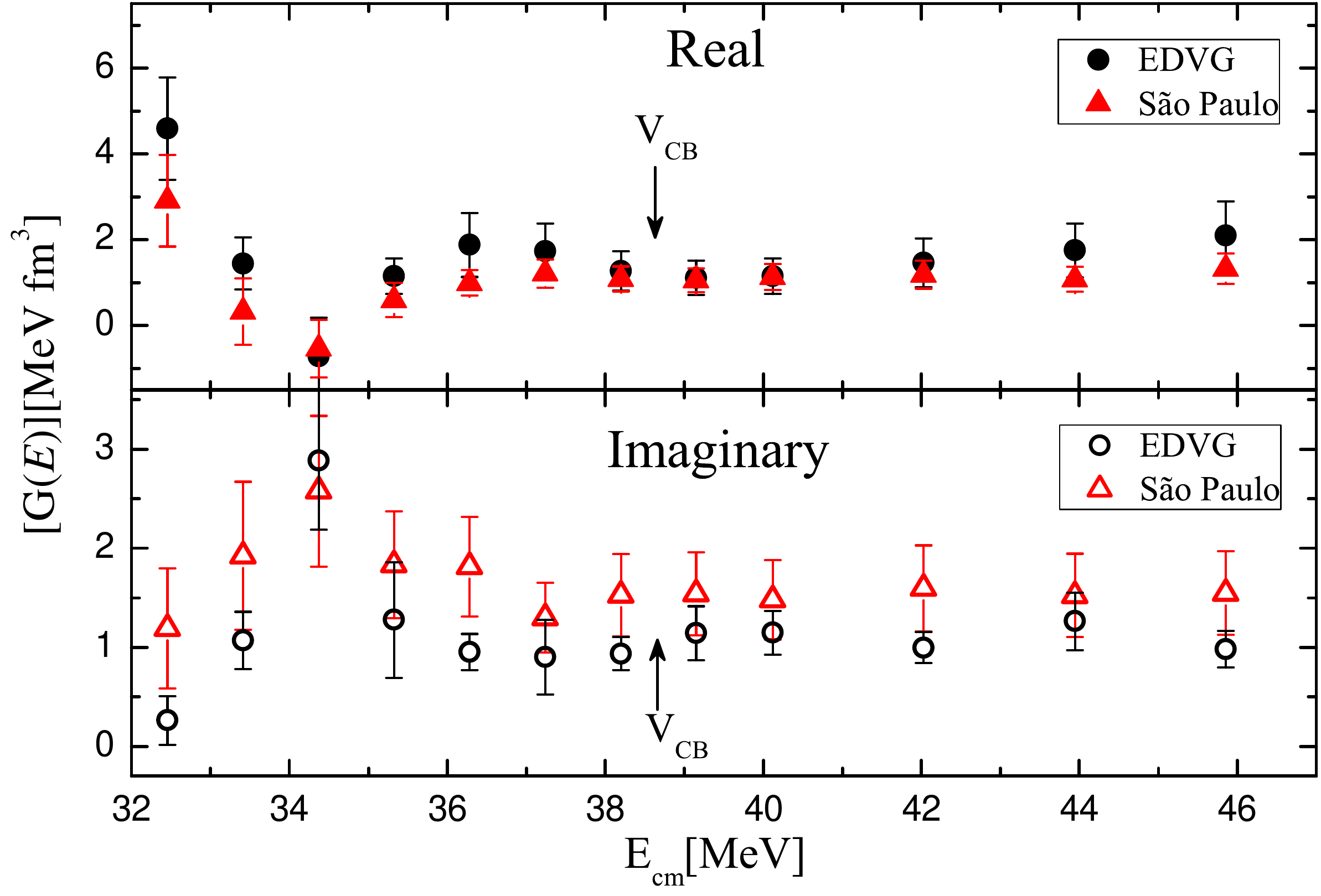}
	\caption{Volume integrals per nucleon $G(E)$ evaluated at the sensitivity radius calculated for the real (full symbols) and imaginary (open symbols) parts of the Woods-Saxon EDVG (black circles) and S\~ao Paulo (red triangles) potentials analyzed in this work. Uncertainty bars include the  component of the sensitivity radio $R_S$. The arrows indicate the value of the Coulomb barrier $V_{CB}$.} 
	\label{fig:Ivol_SP-EDVG}
\end{figure}   

Analogously, Fig.~\ref{fig:Ivol_SP-EDVG} shows the results of the volume integrals obtained by the SPP and Woods-Saxon EDVG potential. It is observed that the energy dependence is the same for the two potentials, with a clear decrease of the real potential as the energy drops below  the Coulomb barrier. The decrease of the real potential is associated with absorptive processes that remove flux from the elastic channel and implies the presence of couplings to non-elastic channels that remain open even at sub-coulombian energies. The couplings are manifested as a real dynamic repulsive polarization potential that diminishes the overall real potential~\cite{Hussein2006}, a behavior compatible with the breakup threshold anomaly (BTA) for weakly bound systems. 
At energies around the coulomb barrier, the strong coupling to the breakup channel could  diminish the real potential. This effect, taken together with the geometrical restrictions of the potentials SPP and Woods-Saxon EDVG could result in a total real potential with negative values, as the one corresponding to $E_{\rm c.m.}=34.4$~MeV and those found in \cite{gollan2018}. However, an effective negative  value for the real potential does not imply a qualitatively different behavior in elastic scattering wave functions nor in the elastic angular distribution.
\begin{figure}
	\centering
	\includegraphics[width=0.65\textwidth]{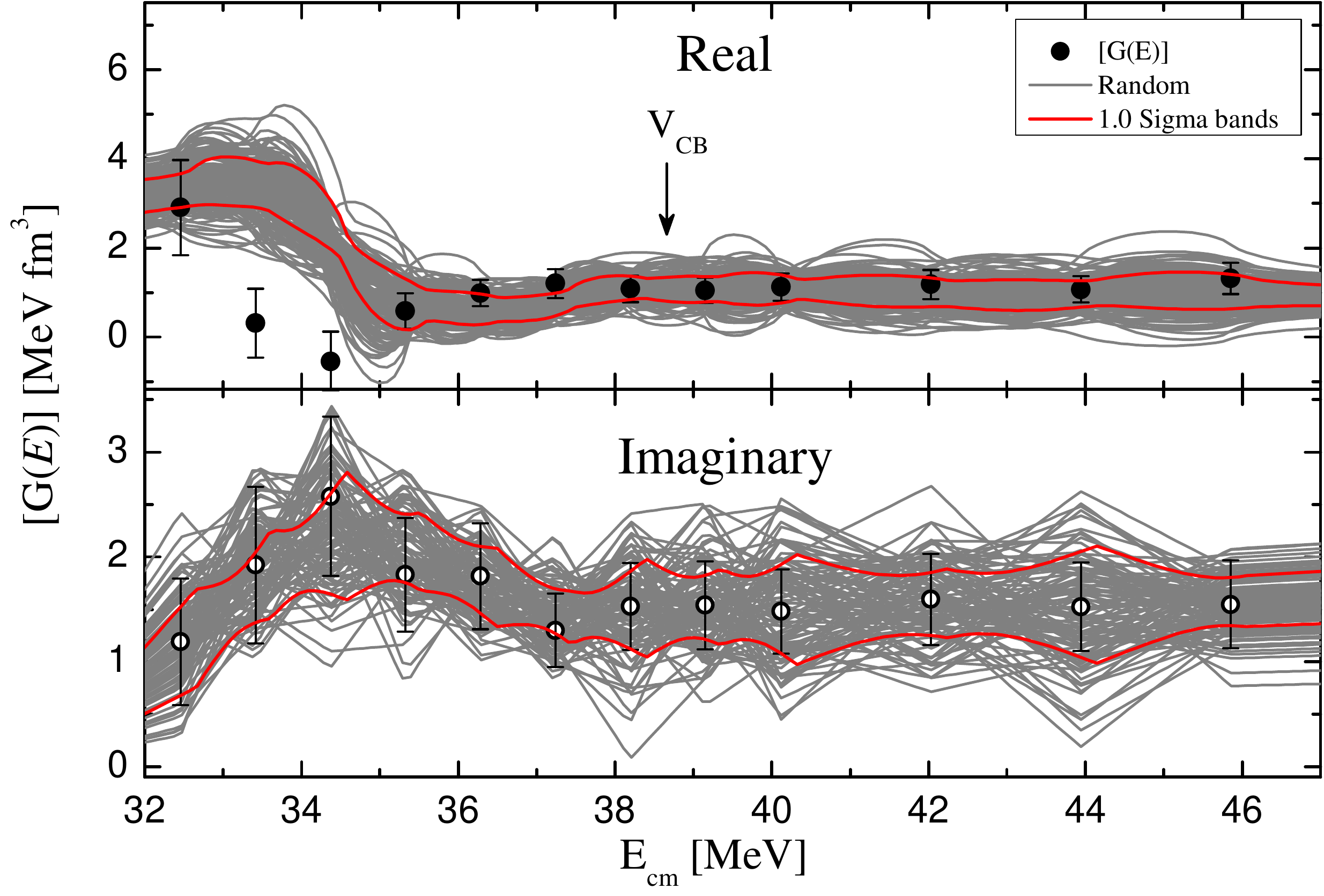}
	\caption{Volume integrals for the real (full circles) and imaginary (open circles) parts of the optical potential calculated for the S\~ao Paulo potential evaluated at $R_S$. The grey lines in the bottom panel represent the schematic segments lines calculated from random points with bivariate normal distribution centered at the adjusted values (200 lines were drawn). The grey lines from the upper graph are the result of the integral of Eq.~\ref{eq:DR_Ivol}. The red lines represent $\pm1\sigma$ confidence bands.}
	\label{fig:DR_random}
\end{figure} 

The next step was to calculate the dispersion relation. Since the energy dependence of the potentials is a direct consequence of the dispersion relation~\cite{satchler1991,Hussein2006}, the real and imaginary potentials (or in this case, the volume integrals per nucleon) must be correlated by an expression like Eq.~\ref{eq:DR_Ivol}. To determine this we implemented the bivariate random sampling method~\cite{gollan2018} in which the simple linear parametrization model~\cite{satchler1991} was replaced by confidence bands obtained from random sampling for each energy, around the adjusted values, following a bivariate normal distribution
\begin{equation}\label{eq:random}
f(\textbf{x})=\frac{1}{\sqrt{(2\pi)^{2}\rm{det}\mathbb{V}}}\exp[{-\frac{1}{2}(\textbf{x} - \textbf{x}_{0})^{\textit{T}}\mathbb{V}^{-1}(\textbf{x} - \textbf{x}_{0})}],
\end{equation}
where $\textbf{x}_{0}=\{[G(E)]_{V0},[G(E)]_{W0}\}$ are the calculated volume integrals per nucleon from the adjusted potential parameters and $\textbf{x}$ are sample values derived from the covariance matrix $\mathbb{V}$~\cite{navarro2014,Abriola2015,abriola2017}. The results for the S\~ao Paulo potential volume integrals are displayed in figure~\ref{fig:DR_random}. 
The grey lines of the real volume integrals represent the results of Eq.~\ref{eq:DR_Ivol} for 200 random lines obtained from the distribution of the imaginary points around the adjusted volume integrals. As it can be seen, all the real volume integrals values lie within the $1\sigma$ confidence bands, with the exception of $E_{\rm c.m.} =$~33.5~MeV and $E_{\rm c.m.} =$~34.4~MeV. This difference may be due to the inherent difficulty in adjusting the potential parameters for energies below the Coulomb barrier, where deviations from pure Rutherford behavior are small, which limits the performance of the numerical codes. Based on this general criterion, we concluded that the relation between the real and imaginary radial integrals, and therefore the potentials, are consistent with the dispersion relation and indicates the presence of the breakup threshold anomaly.

\section{Summary and conclusions}
\label{conclusiones}
In the present work we measured the elastic scattering angular distributions of the weakly bound $^{9}$Be nucleus on a $^{197}$Au target for twelve energies around the Coulomb barrier at the TANDAR laboratory. In addition, we measured the inelastic scattering angular distribution corresponding to two excited states of the target for eleven energies. The elastic scattering angular distributions were analyzed in the framework of the optical model by a phenomenological Woods-Saxon potential and a double folding S\~ao Paulo potential. To analyze the inelastic scattering data, we performed coupled channels calculation with collective model using matrix elements from a particle-asymmetric rotor model considering Coulomb and nuclear deformation potentials and the  obtained Woods-Saxon as a bare potential. The results confirm that the   540(50)~keV peak corresponds to the $3/2^{+}(\rm{g.s.})\rightarrow 7/2^{+}$ transition. Respect to the  274(40)~keV peak, the calculations indicate a contribution from the $3/2^{+}(\rm{g.s.})\rightarrow 5/2^{+}$ transition.

Regarding the energy dependence, the results obtained with both interaction potentials show a peak of the imaginary part below the Coulomb barrier. The  behavior of the real part is exactly the opposite. Through the bivariate random sampling method, we concluded that the energy behavior of both parts are compatible with the dispersion relation. This is a clear indication of absorption due to non-elastic channels that remains open below the Coulomb barrier and of the presence of the breakup threshold anomaly. The low threshold for the fragmentation of the projectile means that coupling to this channel continues to be significant at sub-barrier energies and implies the appearance of  a real dynamic polarization potential that reduce the total real potential.

\section*{Acknowledgements}

This work was partially supported by Consejo Nacional de Investigaciones Cient\'ificas y
Tecnol\'ogicas (CONICET, Argentina) through grant PIP00786CO and Fondo para la
Investigaci\'on Cient\'ifica y Tecnológica (FONCYT, Argentina) through grant PICT-2017-4088.

\end{document}